\documentclass[aps,prb,twocolumn,showpacs,showkeys]{revtex4} 
\usepackage{amsfonts}
\usepackage{bm}
\usepackage{color}
\usepackage{graphicx}
\usepackage{psfrag}

\def\d#1{{\rm d}#1}
\def\diag{{\rm diag}}

\def\gdir#1{\langle #1 \rangle}
\def\gplane#1{\{#1\}}

\def\CRSS{{\rm CRSS}\,}

\def\refeq#1{(\ref{#1})}
\def\reffig#1{Fig.~\ref{#1}}

\begin{document}
\title{Calculation of dislocation positions and curved transition\\ pathways in BCC crystals from atomic displacements}

\author{R. Gr\"oger}
\email{groger@ipm.cz}
\affiliation{Central European Institute of Technology -- Institute of Physics of Materials
  (CEITEC-IPM), Academy of Sciences of the Czech Republic, \v{Z}i\v{z}kova 22, 61662 Brno, Czech
  Republic}

\begin{abstract}
  The thermodynamic description of dislocation glide in crystals depends crucially on the shape of
  the Peierls barrier that the dislocation has to overcome when moving in the lattice. While the
  height of this barrier can be obtained unequivocally using saddle-point search algorithms such as
  the Nudged Elastic Band (NEB) method, its exact shape depends on the chosen approximation of the
  transition pathway of the system. The purpose of this paper is to formulate a procedure that
  allows to identify the position of the dislocation directly from the displacements of atoms in its
  core. We investigate the performance of this model by calculating curved paths of a $1/2[111]$
  screw dislocation in tungsten from a series of images obtained recently using the NEB method at
  zero applied stress and for positive/negative shear stresses perpendicular to the slip
  direction. The Peierls barriers plotted along these curved paths are shown to be quite different
  from those obtained previously by assuming the straight dislocation path.
\end{abstract}

\date{\today}

\pacs{61.72.Hh, 02.60.-x, 45.10.Db}

\keywords{screw dislocation, BCC metal, dislocation pathway, Peierls barrier, Nudged Elastic Band.}

\maketitle


\section{Introduction}

Computational studies of dislocations in body-centered cubic (BCC) metals have been made using a
wide variety of interatomic potentials and ab initio methods \cite{vitek:70, xu:98, woodward:01,
  frederiksen:03, li:04, cawkwell:05, ventelon:07, pizzagalli:08, groger:08a, rodney:09,
  mrovec:11}. They view the evolution of the state of the system as its motion along an a priori
unknown path in the configurational space spanned by the $3N$ atomic degrees of freedom (DOF), where
$N$ is the number of atoms in the simulated block. While this picture arises directly from molecular
simulations, it presents severe difficulties when developing theoretical models of thermally
activated glide of dislocations \cite{dorn:64, groger:08c}, which serve to coarse-grain the
atomistic results to the continuum. Not surprisingly, there has been a long-standing interest in
developing approximate schemes to extract the \emph{effective} position of the dislocation from the
positions of atoms obtained by molecular statics simulations of single dislocations. While this
approach is certainly attractive, it cannot be used without developing a systematic procedure that
maps the $3N$ atomic DOF to the position of the dislocation.

The most obvious choice that leads to the reduction of complexity of the system is to assume that
the dislocation moves between two neighboring lattice sites along the straight line. This is
implicitly assumed in most papers employing the Nudged Elastic Band (NEB) method \cite{jonsson:98,
  henkelman:00b}, where the position of the dislocation along the minimum energy path scales
linearly with the image number \cite{ventelon:07, mrovec:11}. While the obtained barrier can be used
to assess the stability of each intermediate state along the path, it does not constitute the
Peierls barrier that could be used to develop thermodynamic models of dislocation glide such as
that due to Dorn and Rajnak \cite{dorn:64}. Moreover, a straight application of the NEB method to
all DOF in the system leads to nonuniform distributions of dislocation positions
among the images, which affects the shape of the obtained Peierls barrier \cite{groger:12a}. A
significant improvement of these results is obtained using our modification of the NEB method
\cite{groger:12a}, where atomic relaxations are taken into account. This NEB+r method guarantees
that the positions of the dislocation when following the minimum energy path are distributed
uniformly among the images and thus the assumed proportionality between the dislocation position and
the image number is justified. These developments have led to an accurate estimate of the Peierls
barrier of $1/2\gdir{111}$ screw dislocations in BCC W and its changes under the applied stress
\cite{groger:12a, groger:13}. However, these calculations are still based on the assumption that the
path of the dislocation between two neighboring images is a straight line, which is not true in
general.

In principle, it should be possible to deduce the dislocation position (and thus also its path
between two positions in the lattice) directly from the displacements of atoms as obtained from
molecular statics calculations or from the NEB (NEB+r) methods. This idea dates back to Peierls and
Nabarro \cite{peierls:40, nabarro:47}, who associated the dislocation position with the point in the
slip plane at which the displacement parallel to the slip direction, interpolated from the
displacements of atoms, is equal to $b/2$, where $b$ is the magnitude of the Burgers vector of the
dislocation. Since then, this argument was used many times. In particular, it was adopted by
Pizzagalli et al. \cite{pizzagalli:08}, Rodney and Proville \cite{rodney:09} and Proville et
al. \cite{proville:13}, where the position of the dislocation is defined by a single coordinate
corresponding to the distance that the dislocation makes in a well-defined slip plane. The same
level of approximation was also used by Ventelon et al. \cite{ventelon:13} in one of their methods
(disregistry function method) that uses a combination of isotropic elasticity and geometry of the
BCC lattice to define the position of the dislocation.

In general, the movement of the dislocation should be viewed as a three-dimensional event during
which the center of the dislocation transits along a curved path in the vicinity of the slip
plane. In attempt to resolve this path, Ventelon et al. \cite{ventelon:13} also proposed another
model, whereby the dislocation position is identified using a cost function that is based on both
the actual positions of atoms (as obtained from atomistic simulations) and the positions of atoms
obtained from anisotropic elasticity for some trial dislocation position in the $\gplane{111}$
plane. The actual position of the dislocation is then obtained by minimizing this cost function that
is defined as the distance between the two sets of coordinates in the five-dimensional subspace
spanned by the coordinates of the five most displaced atoms around the dislocation. Since the
position of the dislocation is determined by relative displacements of the three atoms closest to
the dislocation in the direction parallel to the Burgers vector, a similar approach can be devised
that is based on inversion of the Eshelby-Stroh sextic formalism (see, for example Hirth and
Lothe \cite{hirth:82}), which provides elastic displacements of atoms corresponding to a given
position of the dislocation. We have investigated this possibility earlier (unpublished work). While
this approximation can be used when the dislocation is in the middle of the lattice site formed by
the nearest three $\gdir{111}$ atomic columns, it quickly worsens as the dislocation gets closer to
any of these columns or the boundary between the neighboring lattice sites.

A different scheme whereby the position of the dislocation is determined by extrapolating
differential displacements between the three atoms surrounding the dislocation into the interior of
this triangle was developed by Itakura et al. \cite{itakura:12}. This approach is closely related to
a purely geometrical concept of barycentric (or trilinear) coordinates known from ternary diagrams
that was originally applied to estimate the position of the dislocation by Heinrich and
Schellenberger \cite{heinrich:71}. While this method makes use of the actual positions of atoms, the
expression of the dislocation position as a linear combination of the displacements of the three
nearest atoms in the direction parallel to the Burgers vector represents a convenient choice that
is, however, not justified physically. This approximation was avoided in the recent work of Dezerald
et al. \cite{dezerald:14} who aimed to reconstruct the two-dimensional Peierls barrier by
interpolating the line energy of the dislocation, calculated by first principles, from two straight
dislocation paths. One connects the neighboring potential minima in the $\gplane{110}$ plane and the
other passes from an atom (``split-core'' configuration) to the so-called ``hard-core'' position in
the $\gdir{110}$ direction perpendicular to the first path. The Peierls potential is then expressed
in the form of a Fourier series with the Fourier coefficients adjusted so as to minimize the least
squares error between the calculated data and the Fourier series. The path of the dislocation
between two neighboring minimum-energy lattice sites in the $\gplane{110}$ plane are obtained using
the disregistry and cost function methods of Ventelon et al. \cite{ventelon:13}. While these paths
are smooth, the corresponding Peierls barriers display sharp maxima for BCC Mo, W, Nb, while this is
somewhat less pronounced for BCC Ta, V and ferromagnetic BCC Fe. This does not agree with the work
of Suzuki et al. \cite{suzuki:95} and our more recent work (Ref.~\onlinecite{groger:07a,
  groger:08c}), which show that in order to reproduce the experimentally measured temperature
dependence of the flow stress, the Peierls potential has to possess a flat maximum.

In this paper, we develop a procedure that provides the position of a $1/2[111]$ screw dislocation
in BCC crystals (and thus the curved dislocation pathway) solely using the actual displacements of
atoms in the dislocation core, without invoking isotropic or anisotropic elasticity. It generalizes
the concept pioneered by Peierls and Nabarro \cite{peierls:40, nabarro:47} in that it considers all
three $\gplane{110}$ planes on which the dislocation can move. These calculations provide three
lines whose intersection defines the position of the dislocation in the perpendicular $(111)$
plane. We demonstrate the performance of this method by calculating the paths of a straight
$1/2[111]$ screw dislocation in BCC W in the three possible $\gplane{110}$ planes from the discrete
snapshots (images) of the system obtained recently using NEB+r calculations \cite{groger:13}. These
calculations are made under zero applied stress and for positive and negative shear stresses
perpendicular to the slip direction. We demonstrate that the shape of the Peierls barrier changes
when considering the curved transition pathway of the dislocation as compared to that obtained
previously by assuming the straight path of the dislocation between two equivalent minimum-energy
configurations in the lattice \cite{groger:13}.


\section{Complexity reduction}

Let us consider that $\bm{X}_i$ is the position of atom $i$ in the perfect lattice. Upon inserting a
$1/2[111]$ screw dislocation, applying additional displacements due to the externally applied
load $\bm{\Sigma}$ and relaxing the atoms, each atom $i$ moves into its new position, $\bm{x}_i^0$,
where $i=1,2,\ldots,N$ are atomic numbers. This configuration corresponds to the stable equilibrium
(a minimum) with energy $E(\bm{x}_1^0\ldots\bm{x}_N^0;\bm{\Sigma})$. In this configuration the
dislocation is at the bottom of the Peierls valley for the applied stress $\bm{\Sigma}$. We will now
consider that the dislocation has moved away from this minimum at constant applied stress
$\bm{\Sigma}$, which is to say that all atoms were displaced from $\bm{x}_i^0$ to $\bm{x}_i$. The
energy of this new (nonequilibrium) configuration will be denoted
$E(\bm{x}_1\ldots\bm{x}_N;\bm{\Sigma})$. The enthalpy of the final state of the system relative to
its initial state per unit length of the dislocation (or, simply, the line enthalpy of the
dislocation) is then
\begin{equation}
  H(\bm{x}_1\ldots\bm{x}_N;\bm{\Sigma}) = \frac{E(\bm{x}_1\ldots\bm{x}_N;\bm{\Sigma}) - 
    E(\bm{x}_1^0\ldots\bm{x}_N^0;\bm{\Sigma})}{l_{dislo}} \ ,
  \label{eq:Hx}
\end{equation}
where $l_{dislo}$ is the length of the dislocation segment contained in the simulated block. 

The left-hand side of \refeq{eq:Hx} seems to imply that the state of the system is described by the
$3N$ DOF associated with the positions of all particles. However, this is rather impractical from
the computational point of view because, in this case, the energy of the system would have to be
calculated from all atoms in the system.

To reduce this complexity, it is more convenient to separate the atomic degrees of freedom into two
classes: (i) the minimum number of DOF that determine the position of the dislocation, and (ii) all
remaining DOF that do not affect the position of the dislocation and thus they represent merely the
large-scale response of the system to incorporating the dislocation. For this purpose, we can use
the concept of the Burgers circuit to identify the region of the atomic block that contains the
dislocation in its interior. When viewing a BCC crystal along the [111] direction, the shortest of
these circuits passes through three atoms that are closest to the center of the dislocation. This
implies that only 9 DOF are necessary to describe the dislocation position, while the remaining
$3N-9$ DOF represent the displacements of other atoms around the dislocation. However, the position
of the dislocation is determined by relative displacements of the three atoms above in the direction
of the Burgers vector. This reduces the 9 DOF to three, which are not all independent because their
differences must sum up to the magnitude of the Burgers vector. Therefore, the information about the
position of the dislocation $(X_D,Y_D)$ is encoded in only two DOF.

The former suggests to find a procedure that maps the positions of atoms, $\bm{x}_1 \ldots \bm{x}_N$, to
the position of the intersection of the dislocation line with the perpendicular $(111)$ plane
\emph{in the perfect lattice} (called hereafter as the dislocation position), i.e.
\begin{equation}
  M : \{\bm{x}_1\ldots\bm{x}_N\} \mapsto (X_D,Y_D) \ .
\end{equation}
It is important to emphasize that the former is defined in the deformed lattice with the
dislocation, while the latter in the perfect lattice. If this mapping exists, the state of the
system can be described using only two variables, the coordinates $(X_D,Y_D)$ of the dislocation in
the $(111)$ plane of the perfect lattice. \footnote{We implicitly assume that all other DOF are
  relaxed upon specifying the dislocation position.} Hence, one can write the line enthalpy of the
dislocation as
\begin{equation}
  H(X_D,Y_D;\bm{\Sigma}) = H(\bm{x}_1\ldots\bm{x}_N;\bm{\Sigma}) \ ,
  \label{eq:HXDYD}
\end{equation}
where the right-hand side is obtained from \refeq{eq:Hx}. Eq.~\refeq{eq:HXDYD} opens the possibility
for developing a model of thermally activated dislocation glide, in which the transformation
of the dislocation core is viewed as a motion of the center of the dislocation in the underlying
Peierls potential.

For further developments, it will be convenient to introduce a transition pathway of the
dislocation, $\xi$, that is defined by a series of discrete points $(X_D,Y_D)$ obtained from
individual snapshots of the system as it moves along the minimum energy path. Hence, the left-hand
side of \refeq{eq:HXDYD} is equivalent to $H(\xi;\bm{\Sigma})$,
\begin{equation}
  H(\xi;\bm{\Sigma}) \equiv H(X_D,Y_D;\bm{\Sigma}) \ ,
  \label{eq:Hxi}
\end{equation}
where $\xi$ represents a particular point $(X_D,Y_D)$ along a curved transition path of the
dislocation. If the dislocation remains a straight line during this transition (which is the case at
0~K), we may express the line enthalpy of the dislocation as 
\begin{equation}
  H(\xi;\bm{\Sigma}) = V(\xi;\bm{\Sigma}^{nonglide}) - \sigma^{glide} b \xi \ , 
\end{equation}
where $V$ is the Peierls barrier \footnote{The minimum of the Peierls barrier
  $V(\xi;\bm{\Sigma}^{nonglide})$ for a given applied stress tensor $\bm{\Sigma}^{nonglide}$ is
  shifted to zero.} and the second term is the work done by the applied stress on displacing the
dislocation by the distance $\xi$ measured along the transition path. In the equation above, we have
split the applied stress as $\bm{\Sigma} = \bm{\Sigma}^{glide} + \bm{\Sigma}^{nonglide}$, where
$\bm{\Sigma}^{glide}$ contains only the shear stress $\sigma^{glide}$ acting in the slip plane
parallel to the slip direction (i.e. the Schmid stress), whereas $\bm{\Sigma}^{nonglide}$ contains
all other stresses. The latter are all stress components that do not exert a Peach-Koehler force on
the dislocation. Clearly, the glide (Schmid) stress does work on displacing the dislocation, while
non-glide stresses affect the shape of the Peierls barrier. The Peierls barrier can then be obtained
from
\begin{equation}
  V(\xi;\bm{\Sigma}^{nonglide}) = H(\xi;\bm{\Sigma}) + \sigma^{glide} b \xi \ .
  \label{eq:Vxi}
\end{equation}
Here, the first term on the right-hand side is obtained directly from the NEB (NEB+r) calculations,
as it is evident by combining \refeq{eq:Hx}, \refeq{eq:HXDYD} and \refeq{eq:Hxi}. The expression
\refeq{eq:Vxi} is completely general and can be used to obtain the shape of the Peierls barrier for
an arbitrary applied load. For the sake of simplicity, we will consider in the following that
$\sigma^{glide}=0$. Hence, the NEB (NEB+r) calculations directly yield the line energies of the
dislocation, $V(\xi;\bm{\Sigma}^{nonglide})$, subject to a given non-glide stress
$\bm{\Sigma}^{nonglide}$. 


\section{Position of the dislocation}
\label{sec:dpos}

\begin{figure}[!htb]
  \centering
  \includegraphics[scale=.65]{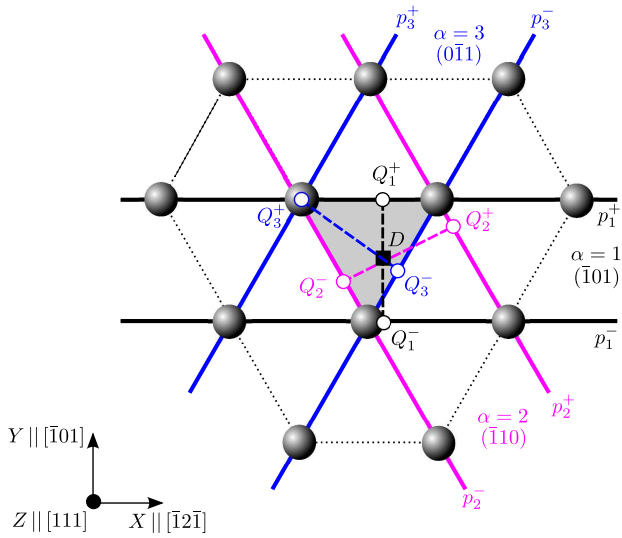}
  \caption{The three possible $\gplane{110}$ slip planes on which the dislocation can move in the
    lattice ($\alpha=1,2,3$). The chains of atoms in the planes immediately above and below the
    lattice site with the dislocation are marked $p_\alpha^\pm$. The dislocation positions within
    these chains are marked $Q_\alpha^\pm$. The sought position of the dislocation corresponds to
    the point $D$, which is defined as an intersection of the three line segments
    $Q_\alpha^+Q_\alpha^-$ for $\alpha=1,2,3$.}
  \label{fig:lattice}
\end{figure}

Upon inserting the dislocation parallel to the $z$ axis, applying external load and relaxing the
atomic positions, the atoms move from their perfect lattice positions, $\bm{X}_i$, into their new
positions, $\bm{x}_i$. These two sets of atomic positions can be used to obtain a differential
displacement map that uniquely identifies the lattice site with the dislocation. One such site
corresponds to the gray triangle in \reffig{fig:lattice} with the dislocation at some unknown
position in this interior. The three $\gplane{110}$ planes on which the dislocation can move are
marked in \reffig{fig:lattice} as $\alpha=1,2,3$ and distinguished by colors.

For each slip plane $\alpha$, we consider the two planes $p_\alpha^\pm$ immediately above and below
the lattice site with the dislocation, which are drawn in \reffig{fig:lattice} by solid
lines. \footnote{The planes above and below the lattice site with the dislocation refer to
  the two planes adjacent to this site. It is not important which of the two planes is considered as
  being ``above'' and which ``below'' this lattice site.} Each of these planes is represented by a
finite number $n_\alpha^\pm$ of atoms $P_{\alpha(i)}^\pm$, where $i=1\ldots n_\alpha^\pm$, which are
ordered such as to form chains. The displacement of each atom projected parallel to the Burgers
vector of the dislocation is then defined as
\begin{equation}
  u_b(P_{\alpha(i)}^\pm) = \frac{\left[ \bm{x}(P_{\alpha(i)}^\pm)-\bm{X}(P_{\alpha(i)}^\pm) \right] \cdot \bm{b}}{b} \ ,
\end{equation}
where $\bm{b}$ is the Burgers vector of the dislocation, and $b$ its magnitude. For the atoms far
away from the dislocation core, the displacements $u_b$ approach constant values that represent the
maximum and minimum of $u_b$ along the chain. Similarly as in Refs.~\onlinecite{peierls:40,
  nabarro:47}, the position of the dislocation in each chain $p_\alpha^\pm$, denoted hereafter
$Q_\alpha^\pm$, is assumed to coincide with the point along the chain, where the displacement in the
direction parallel to the Burgers vector is half-way between its minima and maxima:
\begin{equation}
  u_b(Q_\alpha^\pm) = \frac{1}{2} \left[\min_i u_b(P_{\alpha(i)}^\pm) + \max_i u_b(P_{\alpha(i)}^\pm) \right] \ .
  \label{eq:uavg}
\end{equation}
For each slip plane $\alpha$, this equation thus provides two estimates of the dislocation position,
one for the chain of atoms above ($Q_\alpha^+$) and the other below ($Q_\alpha^-$) the lattice site
with the dislocation. These points then define the line segments $Q_\alpha^+Q_\alpha^-$ that are
plotted in \reffig{fig:lattice} by dashed lines. The actual position of the dislocation is
associated with the intersection of these line segments. This is marked in \reffig{fig:lattice} as
the point $D$. The implicit definition of $Q_\alpha^\pm$ using Eq.~\refeq{eq:uavg} is an
approximation that, nevertheless, provides a good estimate of the position of the dislocation when
it is at the body center of the shaded dislocation triangle in \reffig{fig:lattice}, i.e. when the
system is in equilibrium at zero applied stress. This way of obtaining $Q_\alpha^\pm$ is adopted
here for all states of the system along the minimum energy path irrespective of whether they
represent equilibrium or nonequilibrium states.

\begin{figure}[!htb]
  \centering
  \includegraphics[scale=.45]{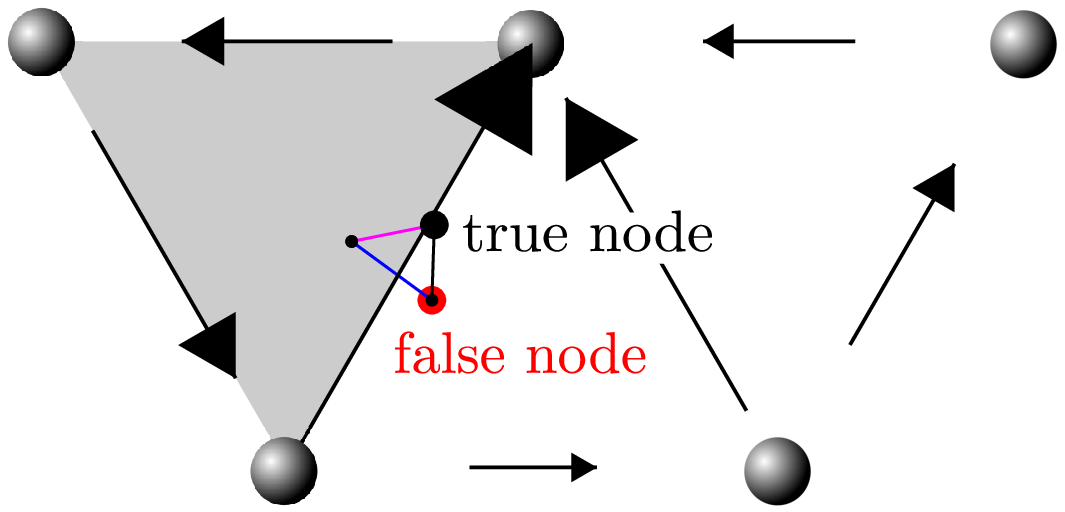} \\
  (a) \\[1em]
  \includegraphics[scale=.4]{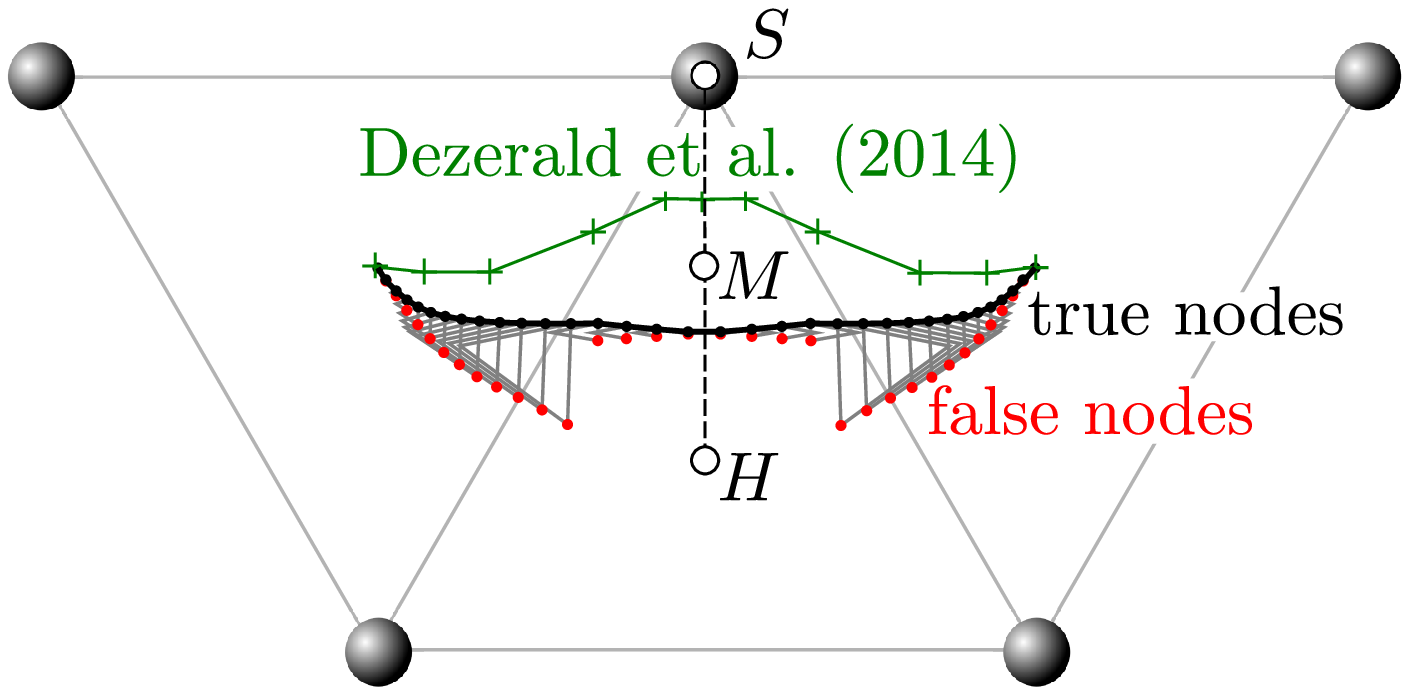} \\
  (b)
  \caption{Approximation of the dislocation position from the image obtained using the NEB+r method,
    where the dislocation is very close to the boundary between two neighboring lattice sites
    (a). The small triangles are estimates of the dislocation position obtained using the method
    described here. The vertex marked ``false node'' [red circle in (a)] lies outside of the gray
    triangle and represents a false prediction of the dislocation position, as described in the
    text. In (b), we show the approximation of the dislocation position along the transition
    path. The black curve marked ``true nodes'' represents the correct approximation of the
    dislocation pathway. For comparison, we plot in green (+ symbols) the path in BCC W obtained by
    Dezerald et al. \cite{dezerald:14} using their cost function method; this is taken from their
    Fig.10(b) and plotted in the orientation of the block shown in \reffig{fig:lattice}.}
  \label{fig:false_node}
\end{figure}

It should be pointed out that there is no a priori guarantee that the three line segments
$Q_\alpha^+Q_\alpha^-$ for $\alpha=1,2,3$ will intersect at a single point. In general, one obtains
three points that correspond to the intersections of three pairs of lines $Q_\alpha^+Q_\alpha^-
\times Q_\beta^+Q_\beta^-$ for the slip planes $\alpha\not=\beta$. This is illustrated in
\reffig{fig:false_node}(a), where the shaded spheres correspond to the positions of atoms in the
perfect lattice. The arrows are relative displacements of atoms between the image along the path for
which the dislocation is closest to the boundary between the two lattice sites in
\reffig{fig:false_node}(a) and the perfect lattice. The three edges of the small triangles are
obtained from the method described above. In particular, the black edge in
\reffig{fig:false_node}(a) lies along $Q_1^+Q_1^-$, purple along $Q_2^+Q_2^-$ and blue along
$Q_3^+Q_3^-$ lines defined in \reffig{fig:lattice}. Each vertex of this small triangle, as well as
any point in its interior, that are within the large gray triangle can be taken as representatives
of the dislocation position. This is, however, not the case for the vertex marked ``false node'',
because it is outside of the gray lattice site determined by the differential displacement
map. \reffig{fig:false_node}(b) shows the transition pathway predicted by the ``true nodes'' (black
dots) and ``false nodes'' (red dots) from a series of images obtained by the NEB+r under zero
applied stress \cite{groger:13}. The two sets of nodes coincide when the system is close to the
beginning, end and the middle of the minimum energy path. For these images, the three line segments
in \reffig{fig:lattice} intersect at a single point and thus the dislocation positions are
determined uniquely.

We define the transition pathway as a curve connecting the ``true nodes'' in
\reffig{fig:false_node}(b). This is quite different from that obtained by the cost function method
employed in Ref.~\onlinecite{dezerald:14} that is plotted by the green line (+ symbols). This
deviation is the largest for the point in the middle of the path, which is determined uniquely by
our method. Dezerald et al. \cite{dezerald:14} also show that the paths of the dislocation at zero
applied stress obtained for various BCC metals are quite similar. This opens the possibility for a
qualitative comparison of the dislocation pathway shown in \reffig{fig:false_node}(b) with Fig.~11
(data marked ``DFT'') and Fig.~14 of Itakura et al. \cite{itakura:12} for BCC Fe. The latter show
that the energy of the dislocation is the lowest when positioned between the points marked $H$
(``hard-core'' position) and $M$ in \reffig{fig:false_node}(b), which agrees well with the path
predicted by our method.

\begin{figure}[!htb]
  \centering
  \includegraphics[scale=.35]{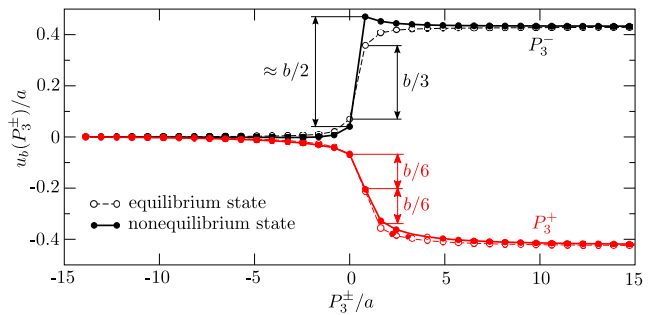}
  \caption{Displacements of atoms parallel to the Burgers vector of the dislocation in the chains
    $p_3^-$ and $p_3^+$ shown in \reffig{fig:lattice} (blue). The dashed curves (``equilibrium
    state'') correspond to the atomic configuration in which the dislocation is at the body center
    of the lattice site (initial image for the NEB calculation at zero applied stress), whereas the
    solid curves (``nonequilibrium state'') are for the image in which the dislocation is close to
    the boundary between the two neighboring lattice sites [see also \reffig{fig:false_node}(a)].}
  \label{fig:ub_pos}
\end{figure}

The origin of the ``false node'' in \reffig{fig:false_node}(a) can be understood by looking at the
displacements $u_b(P_3^\pm)$ of the chain of atoms that are parallel to the boundary between these
neighboring lattice sites (i.e. the blue lines in \reffig{fig:lattice}). These displacements are
shown in \reffig{fig:ub_pos} for two principally different atomic configurations. The first, marked
``equilibrium state'', corresponds to the initial image along the path, where the dislocation is
exactly at the body center of the gray triangle in \reffig{fig:false_node}(a). The second, marked
``nonequilibrium state'', corresponds to the image in which the dislocation is closest to the
boundary with the neighboring lattice site [case shown in \reffig{fig:false_node}(a)]. In the
former, the displacements $u_b(P_3^\pm)$ vary monotonously from their minimum at one end of the
chain to their maximum at the other. In this case, the average displacement of atoms parallel to the
Burgers vector is close to the inflection point of $u_b(P_3^\pm)$. Provided the system is close to
equilibrium, the position of the dislocation along the chain can thus be safely determined using
\refeq{eq:uavg}. For the nonequilibrium state, as the dislocation gets closer to the boundary
between the two neighboring lattice sites, one atomic bond in the chain that coincides with this
boundary becomes stretched to around $b/2$. This is shown in \reffig{fig:ub_pos} and this relative
displacement of atoms is the bigger the closer the dislocation is to the boundary. In this case, the
displacements of atoms $u_b(P_3^-)$ do not vary monotonously along the chain. Instead, the sudden
increase of displacements in the vicinity of the dislocation is followed by a gradual decay of the
displacements, eventually reaching a constant value.

Because the position of the inflection point of $u_b(P_3^-)$ cannot be determined with sufficient
accuracy for nonequilibrium states, the linear approximation \refeq{eq:uavg} may be a poor
representation of the position $Q_3^-$. Due to the nonmonotonous character of $u_b(P_3^-)$ it is
also not advisable to look for higher order schemes. Instead, these observations suggest to make the
prediction of the dislocation position only from two $\gplane{110}$ planes, excluding the one that
is parallel to the boundary between the adjacent lattice sites (in \reffig{fig:false_node}, this is
the case for $\alpha=3$ corresponding to the blue edge of the small triangle). We have also checked
that the jump in $u_b$ vs. $P_3^-$ does not appear if the dislocation position is obtained from
the chain of atoms farther away from the lattice site with the dislocation. However, this leads to
less accurate estimates of the dislocation position (larger triangle obtained by the intersection of
$Q_\alpha^+Q_{\alpha}^-$ with $Q_\beta^+Q_{\beta}^-$ for $\alpha\not=\beta$) and also to the drift
of the dislocation position from the prediction made using the nearest chains of atoms in the three
$\gplane{110}$ planes.

In all calculations made in this paper, the interactions between atoms were described using the Bond
Order Potential (BOP) for BCC tungsten \cite{mrovec:07}. However, we have confirmed that the
nonmonotonous character of the $u_b$ vs. $P_3^-$ curve in \reffig{fig:ub_pos} is not specific to the
potential used. This was found by calculating the dislocation pathway at zero applied stress with
atomic interactions described by the Ackland-Thetford potential \cite{ackland:87} for BCC Ta. This
jump is also not caused by the mismatch betweeen the positions of the far-field atoms that are held
fixed during the NEB (NEB+r) calculations and the actual position of the dislocation
\citep{pizzagalli:08}. This was verified by doubling the radius of the cylindrical block used for
the NEB simulation, which did not produce any noticeable change in the dependence of $u_b$ on
$P_3^{\pm}$. The jump is present in the data obtained from both NEB and NEB+r methods and in all
atomic configurations, where the dislocation is close to the boundary between two adjacent lattice
sites. It is also important to emphasize that the NEB methods do not impose any constraint to keep
the Burgers vector of the dislocation constant during the search for the minimum energy path of the
system. Owing to the above-mentioned misfit of atomic positions around the dislocation and in the
far-field and rounding errors, it is theoretically possible that the relative displacements between
the three atoms closest to the dislocation as obtained from the intermediate images obtained by NEB
(NEB+r) calculations do not sum exactly to $b$. We have checked this possibility for a few
configurations, where the dislocation is very close to the edge of its lattice site. However, no
significant deviation from the Burgers vector being exactly $b$ was detected.

Finally, an important insight is gained by calculating the dislocation positions when applying the
stress directly in molecular statics calculations. In this case, pure shear stress parallel to the
slip direction acting in the $(\bar{1}01)$ plane was applied in steps up to the value just before
the dislocation moved (for details, see Ref.~\onlinecite{groger:08a}). For each relaxed atomic
configuration at different applied stresses, we examined the dependence of $u_b$
vs. $P_3^-$. Interestingly, all curves were smooth without any jump as observed from the NEB (NEB+r)
calculations. Obviously, all atomic configurations obtained from direct molecular statics
calculations are equilibrium states of the system at the given applied stress, while some
intermediate images obtained using the NEB methods correspond to the unstable branch of the Peierls
potential on which the dislocation moves spontaneously to its nearest minimum-energy
configuration. As far as we can judge from our NEB+r calculations using 32 movable images, the
origin of the jump in the $u_b$ vs. $P_3^-$ curve is associated with the inflection point of the
Peierls barrier at which the dislocation can no longer be stabilized by the applied stress.


\section{Numerical simulations}

In our previous publication (Ref.~\onlinecite{groger:13}), we used the NEB+r method
\cite{groger:12a} to investigate how the Peierls barrier of a $1/2[111]$ screw dislocation in BCC
W depends on the applied stress. In these calculations, the dislocation was assumed to move between
two neighboring lattice sites along the straight line, which is a simplification that can now be
removed. Our objective in the following is to use the calculated images that correspond to the
minimum energy path of the system to: (i) identify the position of the dislocation in each image
using the procedure formulated in the previous section, (ii) use these positions to determine the
curved path of the dislocation on the three $\gplane{110}$ planes, and (iii) plot the energies of
these snapshots along these curved paths to obtain the Peierls barriers.

\begin{figure}[!htb]
  \centering
  \includegraphics[scale=.46]{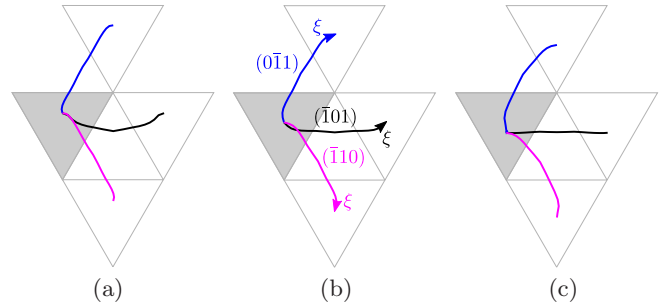} \\
  (a) \hskip2.5cm (b) \hskip2.5cm (c)
  \caption{Dislocation pathways calculated from the positions of atoms in 15 movable images obtained
    using the NEB+r method. The three figures correspond to different shear stresses perpendicular
    to the slip direction: (a) $\tau/C_{44}=-0.04$, (b) $\tau/C_{44}=0$, (c) $\tau/C_{44}=0.04$. The
    calculated dislocation paths are shown by solid lines. The colors distinguish the paths of the
    dislocation on the three $\gplane{110}$ planes: $(\bar{1}01)$ = black, $(0\bar{1}1)$ = blue,
    and $(\bar{1}10)$ = purple. The arrows in the middle panel show the curvilinear coordinates
    $\xi$ for the three paths.}
  \label{fig:dpos}
\end{figure}

We will first investigate how the shape of the dislocation pathway is affected by the shear stress
perpendicular to the slip direction, $\tau$. We carry out these calculations for three
representative values of this stress, namely $\tau/C_{44}=\{-0.04,0,0.04\}$ that is applied by
imposing a uniform stress tensor $\bm{\Sigma} \equiv \bm{\Sigma}^{nonglide} = \diag(-\tau,\tau,0)$
in the coordinate system shown in \reffig{fig:lattice} (for details, see
Ref.~\onlinecite{groger:13}). The calculated paths of the dislocation moving on the three
$\gplane{110}$ planes in the $[111]$ zone and the three values of the shear stresses perpendicular
to the slip direction are shown in \reffig{fig:dpos}. The large triangles represent the boundaries
of four lattice sites -- one from which the dislocation makes the jump (shaded) and the three others
representing the target sites into which the dislocation moves by the glide on the three
$\gplane{110}$ planes. The dislocation paths in bold are obtained as piecewise linear interpolations
of the dislocation positions obtained for each image by the method developed above.

\begin{figure*}[!htb]
  \centering 
  \includegraphics[scale=.46]{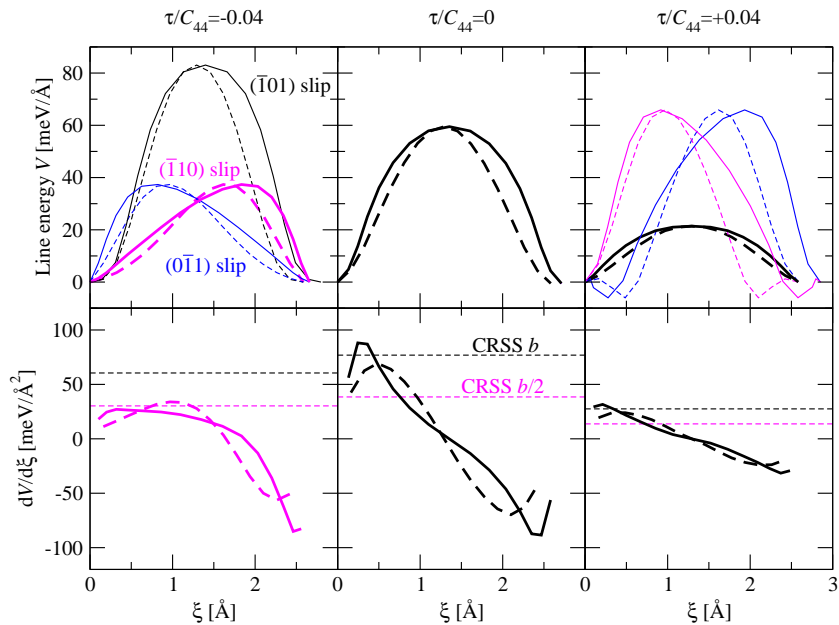}
  \caption{Comparison of the Peierls barriers of the $1/2[111]$ screw dislocation in BCC W
    determined using images obtained from a series of NEB+r calculations \cite{groger:13}. The
    barriers plotted by dashed lines are obtained by assuming that the dislocation moves along the
    straight line connecting the two neighboring potential minima in the glide plane. The solid
    curves are obtained by considering the curved path of the dislocation, as determined by the
    method developed in this paper and shown in \reffig{fig:dpos}. For $\tau=0$ (middle panel), the
    Peierls barriers corresponding to dislocation glide on the three $\gplane{110}$ planes are
    identical. The bottom panel shows the derivatives of the Peierls barriers for the respective
    slip planes.}
  \label{fig:pbarrier}
\end{figure*}

The Peierls barriers for the three $\gplane{110}$ planes and the three values of the shear stress
perpendicular to the slip direction ($\tau$) are shown in the upper panels of
\reffig{fig:pbarrier}. Here, dashed curves correspond to the barriers obtained in
Ref.~\onlinecite{groger:13}, where we considered that the dislocation moves between the two minima
along the straight path. In this case, the transition coordinate is defined as $\xi=a_0I/(M+1)$,
where $I=\{0,1,\ldots,M+1\}$ are image numbers and $a_0=a\sqrt{2/3}$ the distance between
neighboring minimum energy lattice sites in $\gplane{110}$ planes ($a=3.1652~$\AA{}). The solid
curves are obtained by plotting $V(\xi;\bm{\Sigma}^{nonglide})$, obtained from \refeq{eq:Vxi}, along
the paths identified using the procedure developed in Section~\ref{sec:dpos} and shown in
\reffig{fig:dpos}. Firstly, the three barriers for $\tau=0$ are identical, as dictated by symmetry,
which is an important test of our procedure to calculate the dislocation pathway. The shape of the
Peierls barrier changes significantly when drawn along the curved path. In particular, the panels
for zero and positive $\tau$ show that the barriers in black that correspond to the glide of the
dislocation on the $(\bar{1}01)$ plane are wider, with steeper gradient close to $\xi=0$. The
Peierls barrier thus deviates from the $\sin^2\xi$ form to a parabolic or flat-top potentials that
are often used to construct models of thermally activated dislocation glide \cite{suzuki:95,
  groger:08c} (for a review, see the book of Caillard and Martin\cite{caillard:03}). The right panel
in \reffig{fig:pbarrier} shows that the $(0\bar{1}1)$ and $(\bar{1}10)$ barriers develop lower
minima than the initial and final configurations along the path. These are probably consequences of
the fact that the initial and final atomic positions are relaxed down to the maximum force on atom
of only $0.005~$eV/\AA{}.

For completeness, the lower panels of \reffig{fig:pbarrier} show comparisons of the derivatives of
these Peierls barriers. Here, we consider only those barriers that satisfy the fundamental equation
$\CRSS b=\max(\d{V}/\d\xi)$, i.e. the dislocation moves on the $(\bar{1}01)$ plane (black curves,
zero or positive $\tau$), or $\CRSS b/2=\max(\d{V}/\d\xi)$ if it moves on the $(\bar{1}10)$ plane
(purple curves, negative $\tau$). The curvature of the path shifts the peak of $\max(\d{V}/\d\xi)$
towards the beginning of the path. Qualitatively speaking, this shift together with the steep
increase of the Peierls barrier close to $\xi=0$ will affect the activation enthalpy for the
dislocation glide \cite{dorn:64}. This can be demonstrated easily in the limit of zero applied
stress for which the activated segment of the dislocation contains two isolated kinks between
$\xi=0$ and $\xi_{max}$ (the length of the curved transition path of the dislocation). The energy of
this activated state relative to that of the straight dislocation at the bottom of the Peierls
valley is obtained from the Dorn-Rajnak model [Eq.~(6) in Ref.~\onlinecite{groger:08c} with
  $\sigma^*=0$ and $V(\xi_0)=V(0)=0$] and reads
\begin{equation}
  H_b \equiv 2H_k = 2\int_0^{\xi_{max}} \sqrt{[V(\xi)]^2+2EV(\xi)}~{\rm d}\xi \ ,
  \label{eq:Hb}
\end{equation}
where $E=\mu b^2/4 \approx 2.147$~eV/\AA{} is the line tension of a straight
dislocation\cite{groger:08c}.  We now use the two Peierls barriers shown in the middle panel of
\reffig{fig:pbarrier} together with \refeq{eq:Hb} to obtain the predictions of the activation energy
of a pair of noninteracting kinks at zero applied stress \footnote{Since $E \gg V$, the energy in
  \refeq{eq:Hb} can be closely approximated as $2H_k = \sqrt{2E} \int_0^{\xi_{max}}
  \sqrt{V(\xi)}~{\rm d}\xi$.}. These are $2H_k=1.80$~eV if the dislocation is considered to move
along a straight path and $2H_k=2.08$~eV for the curved path of the dislocation [black path in
  \reffig{fig:dpos}(b)]. The latter is in excellent agreement with the value $2H_k=2.06$~eV obtained
in Ref.~\onlinecite{groger:08c} from the experimental data, where the thermal component of the flow
stress vanishes. The curvature of the dislocation pathway and the associated flat maximum of the
Peierls barrier will also give rise to larger curvature of the stress dependence of the activation
enthalpy \cite{dorn:64, groger:08c} and thus to a steep increase of the flow stress with decreasing
temperature, as observed universally in all BCC metals.


\section{Conclusions}

We have developed a numerical scheme that provides the position of the center of the dislocation
only from the knowledge of the displacements of atoms between the relaxed configuration and the
perfect lattice. This represents a generalization of the model devised originally by Peierls and
Nabarro \cite{peierls:40, nabarro:47}. Our model goes beyond these developments in that it provides
two components of the dislocation position, one of which lies in the glide plane, whereas the other
is perpendicular to this plane.

The position of the dislocation in each of the three possible $\gplane{110}$ slip planes is
identified as the point, where the displacement of atoms parallel to the slip direction is half-way
between its minima and maxima. This calculation is made separately for the chain of atoms above and
below the lattice site with the dislocation. For each slip plane, one thus obtains two points that
define a line with possible positions of the dislocation. The pairwise intersections of the three
lines thus obtained define the corners of a triangle that approximates the position of the
dislocation. We have shown that one edge of this triangle is ill-defined when the dislocation is
very close to the boundary between the adjacent lattice site. However, the dislocation path
obtained from the intersections of the two remaining edges is smooth and the dislocation positions
agree with differential displacement maps as well as with DFT calculations made by Itakura et
al. \cite{itakura:12}.

We have demonstrated this procedure by calculating the path of the dislocation between two minimum
energy lattice sites, using the snapshots of atomic positions obtained from our recent NEB+r
calculations made in Ref.~\onlinecite{groger:13}. These calculations have been made for all three
$\gplane{110}$ planes on which the dislocation can move and for zero, positive and negative shear
stresses perpendicular to the slip direction. The curvature of the dislocation path is shown to
affect the shape of the Peierls barrier, which becomes steeper close to the beginning and end of the
transition path and, at the same time, more flat close to its maximum. This shape will affect
macroscopic predictions, such as the temperature dependence of the yield stress. In particular, we
expect that the activation enthalpy will be a stronger function of the applied stress, which will
give rise to a steeper increase of the flow stresses with decreasing temperature. 

\acknowledgments 

The author thanks Vaclav Vitek for fruitful discussions on the topic. This research was supported by
the Marie-Curie International Reintegration Grant No. 247705 ``MesoPhysDef'' and by the Academy of
Sciences of the Czech Republic, Project no. RVO:68081723. This work has been carried out at the
Central European Institute of Technology (CEITEC) with research infrastructure supported by the
project CZ.1.05/1.1.00/02.0068 financed from the EU Structural Funds.

\bibliography{bibliography}
\end{document}